\documentstyle[prl,aps,twocolumn,epsf]{revtex}

\begin{document}
\draft

\twocolumn[\hsize\textwidth%
\columnwidth\hsize\csname@twocolumnfalse\endcsname

\title{\bf Nonuniversal mound formation in nonequilibrium surface growth}

\author{S. Das Sarma, P. Punyindu, and Z. Toroczkai}
\address{Department of Physics, University of Maryland, College Park,
MD 20742-4111}

\date{\today}
\maketitle

\begin{abstract}
We demonstrate, using well-established 
nonequilibrium limited-mobility solid-on-solid
growth models, that mound formation in
the dynamical surface growth morphology does not necessarily 
imply the existence of a surface edge diffusion bias
(``the Schwoebel barrier'').  
We find mounded morphologies in several 
nonequilibrium growth models which incorporate 
{\it no} Schwoebel barrier.
Our numerical results indicate that mounded morphologies in
nonequilibrium surface growth may arise from a number of
distinct physical mechanisms, with the Schwoebel instability
being one of them.
\end{abstract}
\vskip 1pc
Keywords: Computer simulations; Models of surface kinetics;
Molecular beam epitaxy; Scanning tunneling microscopy; Growth;
Surface diffusion; Surface roughening

\vskip 1pc]
\narrowtext

In vacuum deposition growth of thin films or epitaxial layers 
(e.g. MBE) it is common \cite{1} to find mound
formation in the evolving dynamical surface growth morphology.
Although the details of the mounded morphology could differ
considerably depending on the systems and growth conditions,
the basic mounding phenomenon in surface growth has been
reported in a large number of recent experimental publications
\cite{1}.
The typical experiment \cite{1} monitors vacuum deposition
growth on substrates using STM and/or AFM spectroscopies.
Growth mounds are observed
under typical MBE-type growth conditions, and the resultant 
mounded morphology is statistically analyzed 
by studying the dynamical surface
height $h({\bf r},t)$ as a function of the position
${\bf r}$ on the surface and growth time $t$.
Much attention has focused on this ubiquitous phenomenon
of mounding and the associated pattern formation during 
nonequilibrium surface growth for reasons of possible
technological interest (e.g. the possibility of producing 
controlled nanoscale thin film or interface patterns)
and fundamental interest (e.g. understanding nonequilibrium 
growth and pattern formation).

The theoretical interpretation of the mounding phenomenon has 
often been based \cite{1}
on the step-edge diffusion bias \cite{2} or the so-called
Schwoebel barrier \cite{3} effect
(also known as the Ehrlich-Schwoebel \cite{3},
or ES, barrier).
The basic idea of the ES barrier-induced mounding 
(often referred to as an instability) 
is simple :
The ES effect produces an additional energy barrier for
diffusing adatoms on terraces from coming ``down'' toward
the substrate, thus probablistically inhibiting attachment
of atoms to lower or down-steps and enhancing their attachment
to upper or up-steps; the result is therefore mound formation
because deposited atoms cannot come down from upper to lower
terraces and so three-dimensional mounds or pyramids result
as atoms are deposited on the top of already existing terraces.

The physical picture underlying mounded growth under an ES
barrier is manifestly obvious, and clearly the existence of
an ES barrier is a {\it sufficient condition} \cite{2} for
mound formation in nonequilibrium surface growth. Our interest
in this paper is to discuss the {\it necessary} condition for
mound formation in nonequilibrium surface growth morphology
--- more precisely, we want to ask the inverse question, namely,
whether the observation of mound formation  
requires the existence of an ES barrier.
Through concrete examples we
demonstrate that the mound formation in 
nonequilibrium surface growth morphology does {\it not}
necessarily imply the existence of an ES barrier, and we contend
that the recent experimental observations of mound formation in
nonequilibrium surface growth morphology should not be taken as
definitive evidence in favor of an ES barrier-induced
universal mechanism for pattern formation in surface growth.
Mound formation in nonequilibrium surface growth is 
a non-universal phenomenon, and could have very different 
underlying causes in different systems and situations,
with the Schwoebel instability being one particular
mechanism (among many) for the mounded morphology.

Before presenting our results we point out that the possible 
nonuniversality in surface growth mound formation 
(i.e. mounds do not necessarily imply an ES barrier)
has recently been mentioned in at least two experimental
publications \cite{4,5} where it was emphasized that the 
mounded patterns seen on Si \cite{4} and
GaAs \cite{5}, InP \cite{5} surfaces during MBE growth
were not consistent with the phenomenology of a Schwoebel instability.
In two other recent experimental publications \cite{6}
mound formation during semiconductor surface growth
(Ge, GaAs) was carefully analyzed using the prevailing
Schwoebel instability phenomenology with a conclusion
not very dissimilar from that in ref. \cite{5}.
In particular, the ES barrier-based analyses of the 
experimental data in both the papers in refs. \cite{6}
produced rather weak Schwoebel effects in both experiments,
leading to the conclusion in both experiments that the
Schwoebel instability in all likelihood is playing a
small to negligible role in the observed mound formations
in refs. \cite{6}.
Very recently, experimental observations of striking mound
formation \cite{6p} in Au and MgO vapor deposition growth 
have been interpreted without invoking any ES barrier effect.
Thus, the observed mound formation in the nonequilibrium 
surface growth in refs. \cite{4,5,6,6p} is interpreted 
essentially without invoking any key role being played by an ES barrier
whereas the mounded growth morphologies in refs. \cite{1}
have mostly been interpreted as arising essentially due
to a Schwoebel instability.
Thus the inevitable conclusion from recent experimental 
observations \cite{4,5,6,6p} is that the mound formation
in surface growth does not necessarily arise from the
universal mechanism of a Schwoebel instability \cite{1},
but may be caused by different non-universal mechanisms
in different experimental situations.
The purpose of the current article is to explore
this nonuniversality in the mound formation in some
detail using simple solid-on-solid (SOS) growth models
where the kinetic mechanisms leading to the mounded
morphologies are explicitly obvious, and therefore
compelling conclusions can be drawn about the precise
physical mechanism producing the mounds.
A direct comparison between experimental results and
our rather simplistic limited mobility nonequilibrium SOS
models, however, is unwarranted due to the extreme 
simplicity in the growth and diffusion rules in
our models --- our models do serve the purpose of 
explicitly demonstrating the fact that mounded morphologies
can arise without any ES barriers whatsoever.

There have been two proposed mechanisms 
in the literature which lead to mounding without any explicit
ES barrier: One of them invokes \cite{7} a preferential
attachment to up-steps compared with down-steps
(the so-called ``step-adatom'' attraction),
which, in effect, is equivalent to having an ES barrier
because the attachment probability to down-steps is
lower than that to up-steps exactly as it is in the
regular ES barrier case \cite{2,3} --- we therefore do not
distinguish it from the ES barrier mechanism, and in fact,
within the simple growth models we study, these
two {\it energetic} mechanisms 
are physically and mathematically indistinguishable.
The second mounding alternative \cite{8},
which is a purely topologic-kinetic effect,
is the so-called {\it edge diffusion} induced mounding,
where diffusion of adatoms
around cluster edges is shown to lead to mound
formation during nonequilibrium surface growth even in the
absence of any finite ES barrier. One of the concrete examples
we discuss below, the spectacular pyramidal pattern formation
(Fig. 3(c)) in the 2+1 dimensional (d) noise reduced
Wolf-Villain (WV) model \cite{9}, arises from such a
nonequilibrium edge diffusion effect (perhaps in a somewhat
unexpected context).
We also demonstrate, using the WV model and the
Das Sarma-Tamborenea (DT) model \cite{10}, that mound
formation during nonequilibrium surface growth is, in fact,
almost a generic feature of {\it limited mobility} 
solid-on-solid discrete growth
models \cite{9,10,11}, which typically have comparatively
large values of the roughness exponent \cite{11} ($\alpha$)
characterizing the growth morphology.
We find that a large roughness exponent coupled with
atomistic solid-on-solid growth almost invariably
leads to visually mounded growth morphology.
Below we demonstrate that mound formation
in surface morphology arising from this generic
``large $\alpha$'' effect (without any explicit ES barrier)
is often qualitatively virtually indistinguishable from that in 
growth under an ES barrier.
Mound formation in the presence of strong edge diffusion
\cite{8} (as in the d=2+1 WV model in Fig. 3)
is, on the other hand, morphologically quite distinct from the
ES barrier- or the large $\alpha$- induced mound formation.

Our results are based on the extensively studied \cite{11} limited
mobility SOS
nonequilibrium WV \cite{9} and DT \cite{10} growth models.
Both models have been widely studied \cite{11}
in the context of kinetic surface roughening in nonequilibrium
solid-on-solid epitaxial growth --- the interest in and the
importance of these models lie in the fact that these were the
first concrete physically motivated growth models falling
outside the well-known Edwards-Wilkinson-Kardar-Parisi-Zhang \cite{11}
generic universality class in kinetic surface roughening.
Both models involve random deposition of atoms on a square lattice
singular substrate (with a growth rate of 1 layer/sec.
where the growth rate defines the unit of time) under the
SOS constraint with no evaporation or desorption.
An incident atom can diffuse instantaneously before incorporation
if it satisfies certain diffusion rules which differ slightly
in the two models.
In the WV model the incident atom can diffuse within a
diffusion length $l$ (which is taken to be one with the
lattice constant being chosen as the length unit, i.e. only
nearest-neighbor diffusion, in all the results shown in this
paper --- larger values of $l$ do not change our conclusions)
in order to maximize its local coordination number or 
equivalently the number of nearest neighbor bonds it forms
with other atoms (if there are several possible final sites 
satisfying the maximum coordination condition equivalently
then the incident atom chooses one of those sites with equal
random probability and if no other site increases the local
coordination compared with the incident site then the 
atom stays at the incident site).
The DT model is similar to the WV model except for two 
crucial differences:
(1) only incident atoms with no lateral 
bonds (i.e. with the local coordination number of one ---
a nearest-neighbor bond to the atom below is necessary
to satisfy the SOS constraint) are allowed
to diffuse (all other deposited atoms, with one or more 
lateral bonds, are incorporated into the
growing film at their incident sites);
(2) the incident atoms move only to {\it increase} their 
local coordination number (and {\it not} to maximize it
as in the WV model) --- all possible incorporation sites
with finite lateral local coordination numbers are accepted
with random equal probability.
Although these two differences between the DT and the WV model
have turned out to be crucial in distinguishing their
{\it asymptotic} universality class, the two models
exhibit very similar growth behavior for a long transient
pre-asymptotic regime.
It is easy to incorporate \cite{12} an ES barrier in the
DT (or WV) model by introducing differential probabilities
$P_u$ and $P_l$ for adatom attachment to an upper and a lower 
step respectively --- the original DT model \cite{10} has
$P_u=P_l$, and an ES barrier can be explicitly 
incorporated \cite{12} in the model by having 
$P_l < P_u \leq 1$. We call this situation \cite{12}
the DT-ES model (we use $P_u = 1$ throughout with no
loss of generality). We also note, as mentioned above,
that within the DT-ES model the ES barrier 
\cite{2,3} ($P_l < P_u$) and the step-adatom attraction
\cite{7} ($P_u > P_l$) are manifestly equivalent, 
and we therefore do not consider them as separate mechanisms.
We note also that in some of our simulations below we 
have used the noise reduction technique \cite{12,13} 
which have earlier been successful in limited mobility growth
models in reducing the strong stochastic noise effect
through an effective coarse-graining procedure.
All three models described above are studied in both
one-dimensional substrate (d=1+1) and two-dimensional
substrate (d=2+1) systems with periodic boundary conditions
being used in all simulations.
Detailed descriptions of DT and WV models are available
in the literature \cite{9,10,11,12,13}.

In Fig. 1 and 2 we present our d=1+1 
growth simulations, which demonstrate the point
we want to make in this paper.  
We show in Fig. 1 the simulated growth morphologies at three
different times for four different situations, two of which
(Fig. 1(a),(b)) have finite ES barriers and the other
two (Fig. 1(c),(d)) do not. 
\begin{figure}[htbp]
\hspace*{-1cm}
\epsfxsize=3.6 in
\epsfbox{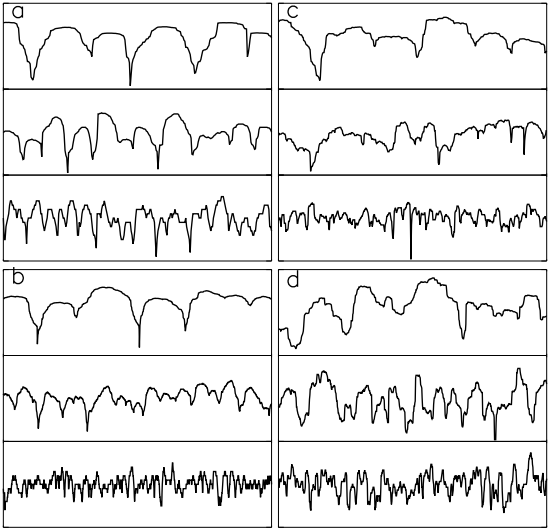}
\vspace*{-3.5cm}
\caption{
Dynamical morphologies at $10^2$, $10^4$ and $10^6$ monolayers (ML)
for (a) DT-ES with $P_l=0.5$, $P_u=1$;
(b) DT-ES with $P_l=0.9$, $P_u=1$;
(c) DT; and
(d) WV models.
}
\end{figure}
\noindent
The important point we wish to emphasize is that, while the
four morphologies and their dynamical evolutions shown in 
Fig. 1 are quite distinct in their details, they all share one
crucial common feature: they all indicate mound formation
although the details of the mounded morphologies 
and the controlling length
scales are obviously quite different in the different cases.
Just the mere observation of mounded morphology, which is 
present in Figs. 1(c),(d), thus does not necessarily 
imply the existence of an ES barrier. 
To further quantify the
mounding apparent in the simulated morphologies of Fig. 1
we show in Fig. 2 the calculated height-height correlation
function, 
$H(r) \sim \langle h({\bf x}) h({\bf r} + {\bf x})
           \rangle_{\bf x}$,
along the surface for two different times.
\begin{figure}[htbp]
\hspace*{-1cm}
\epsfxsize=3.6 in
\epsfbox{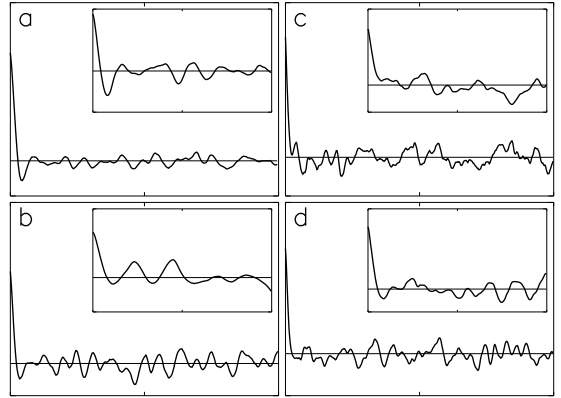}
\vspace*{-2.5cm}
\caption{
The height-height correlation function $H(r)$ at $10^2$ML
(main plots) and $10^4$ML (insets) corresponding respectively to the
morphologies in Fig. 1.
}
\end{figure}
\noindent
All the calculated $H(r)$ show noisy oscillations as a function
of $r$, which implies mound formation
(corresponding to the noisy mounded morphologies of Fig. 1).
It is indeed true that the presence of considerable stochastic
noise associated with the deposition process in the DT, WV models
make the $H(r)$-oscillations quite noisy, but 
the important feature to note in Fig. 2 is that the 
qualitative oscillatory nature of $H(r)$
in situations with or without an ES barrier is essentially the same.
Thus, the mound formation, although noisy, is qualitatively
similar with or without an ES barrier in Figs. 1 and 2.
We have explicitly verified that such growth mounds (or
equivalently $H(r)$ oscillations) are completely absent in the growth 
models \cite{11} which correspond to the generic
Edwards-Wilkinson-Kardar-Parisi-Zhang universality class,
and arise only in the DT, WV limited mobility growth models
which are known to have large value of
the roughness exponent $\alpha$
arising from (linear or nonlinear) surface diffusion processes.
In fact, the effective $\alpha$ in the DT, WV models
is essentially \cite{9,10,11} unity (in d=1+1),
which is the same as what one expects in a naive theoretical
description of growth under the ES barrier
(although the underlying growth mechanisms are completely
different in the two situations).
We believe that any surface growth involving a 
``large'' roughness exponent
( $0.5 < \alpha \lesssim 1$)
will invariably show ``mounded'' morphology
independent of whether there is an ES barrier in the
system or not. We contend that this 
effectively large $\alpha$
is the physical origin for the mounded morphology in 
semiconductor MBE growth where one expects the
surface diffusion driven
linear or nonlinear conserved fourth order
(in contrast to the generic second order) dynamical growth
universality \cite{11} class 
to apply which has the asymptotic exponent :
$\alpha$ (d=1+1) $\approx$ 1; 
$\alpha$ (d=2+1) $\approx$ 0.67 (nonlinear), 1 (linear).
One recent experimental paper \cite{5}, which reports
the observation of mounded GaAs and InP growth with
$\alpha \approx 0.5 - 0.6$, has explicitly made this case,
and other recent reported mound formations \cite{6}
in semiconductor MBE growth are also consistent with our 
contention that mounds may arise from a large
effective roughness exponent
rather than a Schwoebel instability.
Two very recent experimental publications
\cite{6p} have reached the same conclusion in non-semiconductor
MBE growth studies --- in these recent publications \cite{6p}
spectacular mounded surface growth morphologies have been 
interpreted on the basis of the fourth order 
conserved growth equations \cite{9,10,11,12,13}. 
The crucial message of our simulated d=1+1 growth morphologies
in Fig. 1 and 2 is the fact that the mound formation
with [Figs. 1 (a),(b) and 2 (a),(b)] and without ES barrier
[Figs. 1 (c),(d) and 2 (c),(d)] are qualitatively similar,
and therefore the mere observation of a mounded morphology
does not necessarily imply a Schwoebel instability.

\begin{figure}[htbp]
\hspace*{1cm}
\epsfxsize=2.1 in
\epsfbox{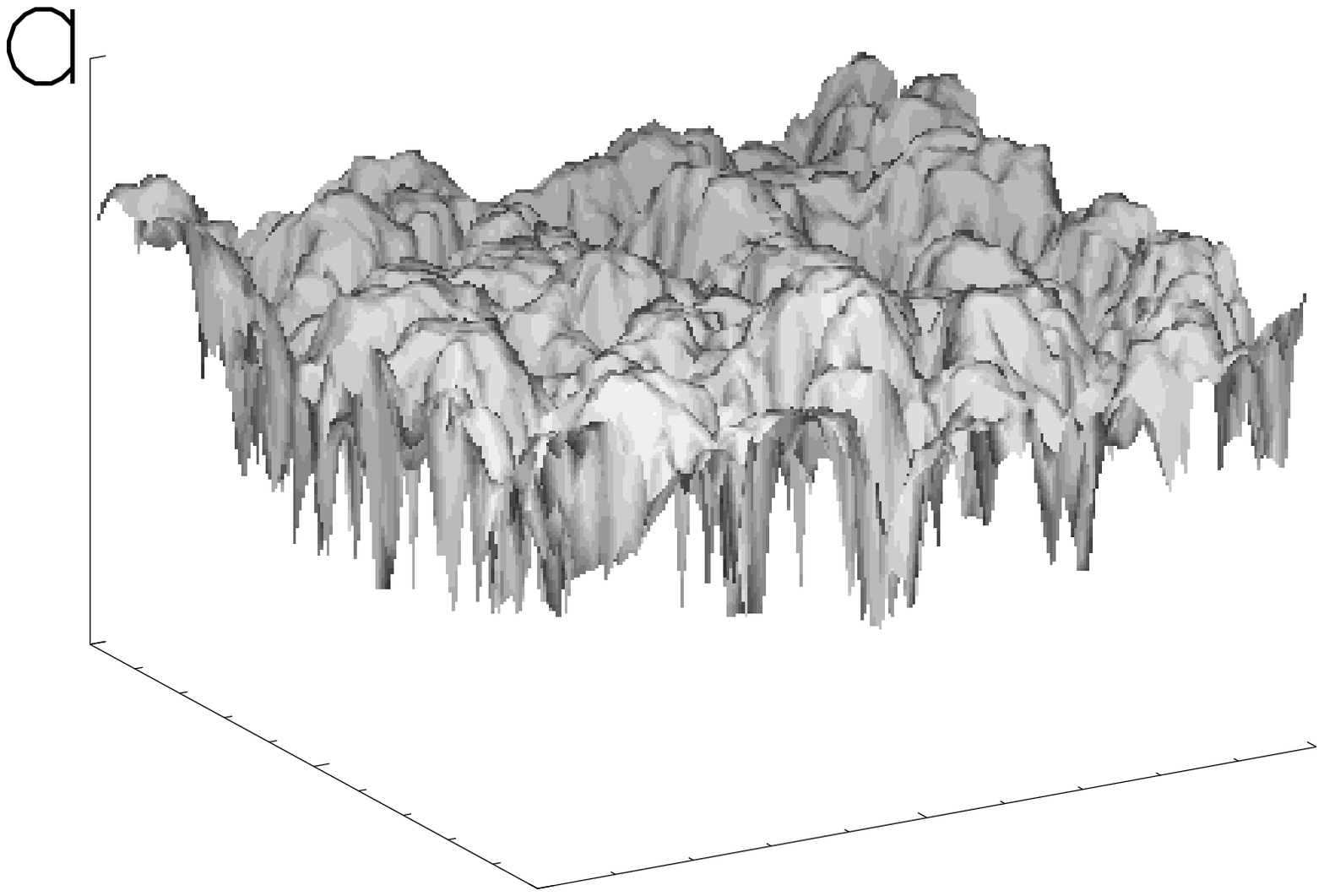}
\vspace*{-1.1cm}
\end{figure}
\begin{figure}[htbp]
\hspace*{1cm}
\epsfxsize=2.1 in
\epsfbox{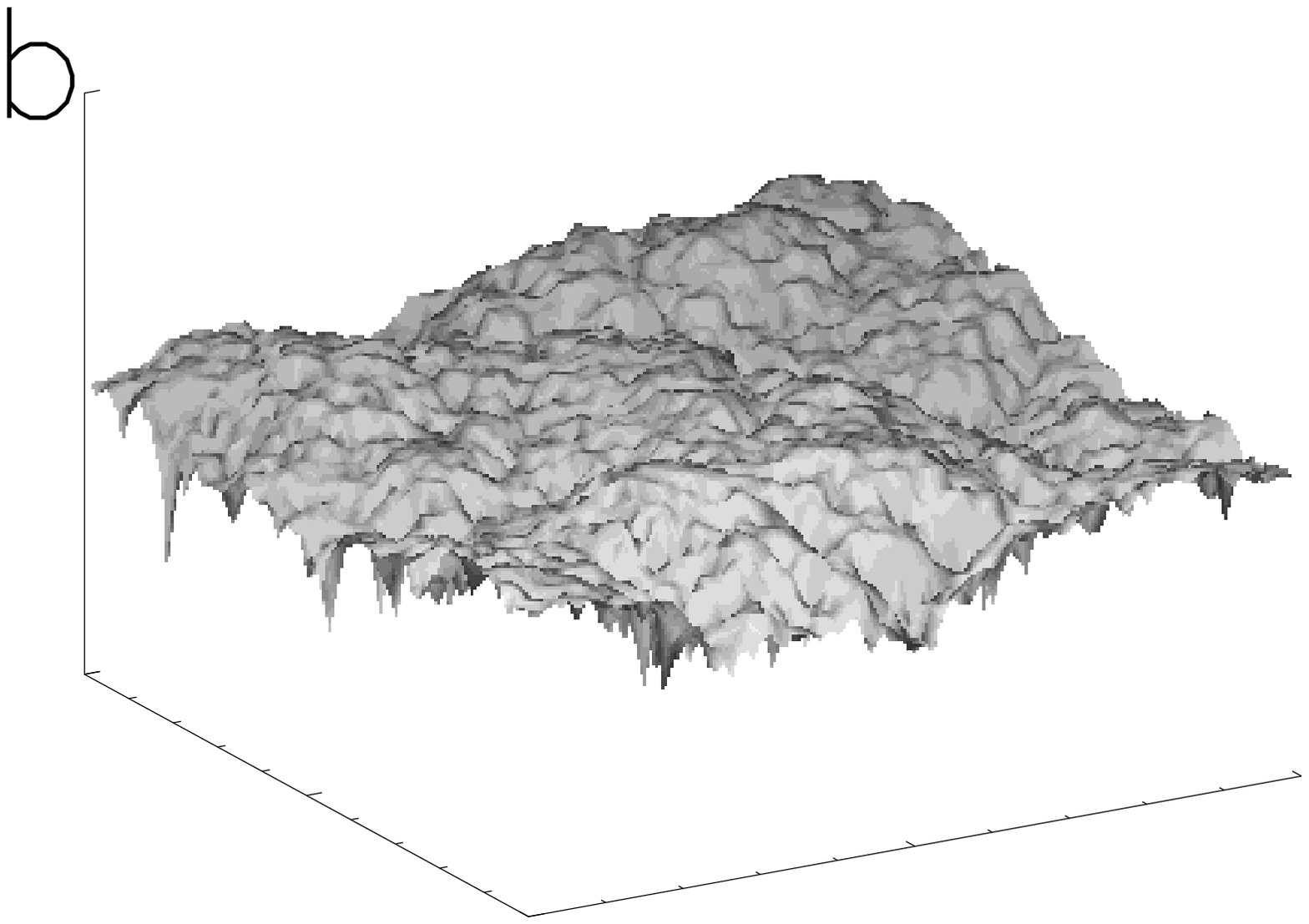}
\vspace*{-1.1cm}
\end{figure}
\begin{figure}[htbp]
\hspace*{1cm}
\epsfxsize=2.1 in
\epsfbox{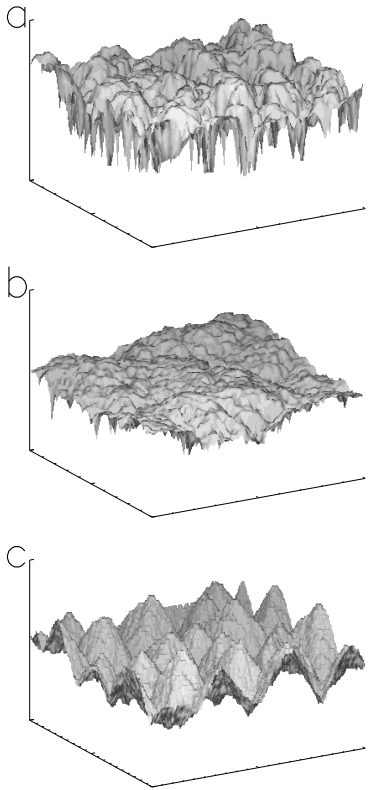}
\caption{
Morphologies from the (a) DT-ES with $P_l=0.5$, $P_u=1$;
(b) DT; and (c) noise reduced WV models.
}
\end{figure}
Finally, in Figs. 3 and 4 we present our results for
the physically more relevant d=2+1 nonequilibrium surface growth.
In Fig. 3(a)-(c) we show the growth morphologies for
the DT-ES, DT,
and the noise-reduced WV model, 
respectively whereas in Fig. 4 we show the scaled
height-height correlation function for the mounded morphologies
depicted in Fig. 3. 
It is apparent that all three models 
(one with an ES barrier and the other two without)
have qualitatively similar oscillations in $H(r)$
indicating mounded growth, and the differences 
in the mounding between
the growth models are purely quantitative.
Again, the important point is that mounded morphologies 
with and without ES barriers manifest similar oscillating
in $H(r)$, indicating that such an oscillatory 
height-height correlation function by itself does not
establish a Schwoebel instability.
Thus we come to the same conclusion: 
mound formation, by itself, does not imply the 
existence of an ES barrier; 
the details of the morphology 
obviously will depend on the existence (or not)
of an ES barrier. 
\begin{figure}[htbp]
\hspace*{0.5cm}
\epsfxsize=3 in
\epsfbox{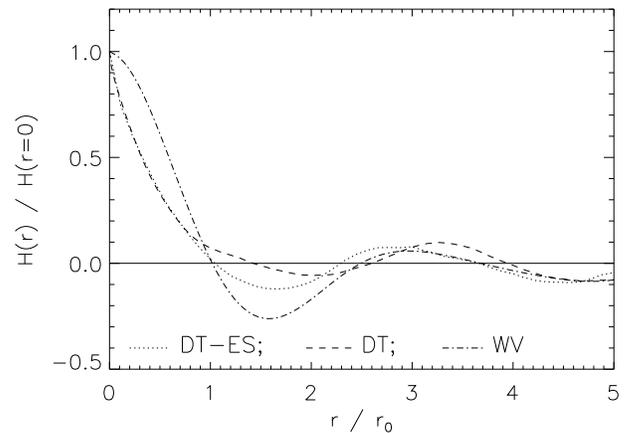}
\vspace*{0.5cm}
\caption{
The scaled $H(r)$ correlation functions corresponding to the
morphologies shown in Fig. 3. ($r_0 \equiv$ mound radius.)
}
\end{figure}
\noindent
We note that the effective values
of the roughness exponent are very similar in 
Fig. 3(a) and (b) i.e. with and without
an ES barrier, both being approximately
$\alpha \sim 0.5$
(far below the asymptotic value $\alpha \approx 1$
expected in the ES barrier growth --- we have verified that
this asymptotic $\alpha \approx 1$ is achieved in our 
simulations at an astronomically long time of $10^9$ layers).
The most astonishing result we show in Fig. 3 is the
spectacular pyramidal mound formation in the
d=2+1 noise reduced WV model (without any
ES barrier), which has not earlier been reported in the literature.
The strikingly regular pyramidal pattern
formation (Fig. 3(c)) in our noise reduced WV model
in fact has a magic slope and
strong coarsening behavior.
The pattern is very reminiscent of the theoretical
growth model studied earlier in ref. \cite{15} in the context of
nonequilibrium growth under an ES barrier where very
similar patterns with slope selection were proposed as
a generic scenario for growth under a Schwoebel instability.
In our case of the noise reduced d=2+1 WV model 
of Fig. 3(c), there is no ES barrier, but there is strong
cluster-edge diffusion.  
This strong edge
diffusion (which obviously cannot happen in 1+1 dimensional growth)
arises in the WV model (but {\it not} in the DT model)
from the hopping of adatoms which have finite lateral 
nearest neighbor bonds (and are therefore the edge
atoms in a cluster).
This edge diffusion 
(discussed in entirely different contexts in \cite{8})
leads to an ``uphill'' surface current
in the 111 direction,
which leads to the formation of the slope-selected 
pyramidal patterned growth morphology.
While noise reduction enhances the edge current
strengthening the pattern formation
(the uphill current is extremely weak in the 
ordinary WV model due to the strong suppression by
the deposition shot noise),
our results of Fig. 3 estabish compellingly
that the WV model in d=2+1 is, in fact,
unstable (uphill current) in contrast to the situation
in d=1+1.  
Thus, the WV model belongs to 
totally different universality classes in d=1+1 and 2+1
dimensions!
We mention that in (unphysical) higher (e.g. d=3+1, 4+1, etc.)
dimensions, the WV model would be even more unstable, 
forming even stronger mounds since the edge diffusion
effects will increase substantially in higher dimensions
due to the possibility of many more configurations of
nearest-neighbor bonding. 
Such unstable growth in high dimensional (d $>$ 2+1)
WV model has earlier been reported \cite{16}
in the literature without any physical explanation.
We have therefore provided the explanation for the
long-standing puzzle of an instability in high-dimensional
(d $>$ 2+1) WV model simulations which were reported
\cite{16} in the literature some years ago.
More details on this phenomenon will be published elsewhere \cite{17}.

In conclusion, we have shown through concrete examples
that, while a
Schwoebel instability is certainly sufficient to cause 
mounded surface growth morphology, the reverse
is not true :
an ES barrier is by no means necessary to produce mounds, 
and mound formation in nonequilibrium surface growth morphology 
does not necessarily imply the existence of a Schwoebel
instability. In particular, we show that a large roughness
exponent (without any ES barrier) as in the fourth order
conserved growth universality class \cite{9,10,11,12,13}
produces mounded growth morphologies which are
indistinguishable from the ES barrier effect.
Any experimentally observed mounded morphology therefore
requires a careful and detailed quantitative analyses
\cite{5,6,6p} to determine the physical mechanism
(e.g. ES barrier, edge diffusion, large roughness exponent
without any ES barrier) underlying its cause
--- in particular, the existence of a mounded growth
morphology by itself may not imply the existence of
any significant Schwoebel barrier.

This work is supported by the NSF-DMR-MRSEC and the US-ONR.

\end{document}